\documentclass{aastex}
\usepackage{natbib}
\usepackage{verbatim}
\usepackage{epsfig}
\def\re{{\rm Re}}
\def\im{{\rm Im}}
\def\kms{{\rm km\ s}^{-1}}
\def\gtwid{\mathrel{\raise.3ex\hbox{$>$\kern-.75em\lower1ex\hbox{$\sim$}}}}
\def\ltwid{\mathrel{\raise.3ex\hbox{$<$\kern-.75em\lower1ex\hbox{$\sim$}}}}
\def\haf{{\textstyle{{1}\over{2}}}}

\def\pmb#1{\setbox0=\hbox{#1}%
  \kern-.025em\copy0\kern-\wd0
  \kern.05em\copy0\kern-\wd0
to  \kern-.025em\raise.0433em\box0 }
\def\bx{{\bf x}}

\shortauthors{GWINN}
\shorttitle{Scintillation from HI}
\begin{document}
\input epsf

\title{\bf Small-Scale Variations of HI Spectra from Interstellar Scintillation}
\author{C.R. Gwinn}
\affil{Physics Department, University of California,
Santa Barbara, California, 93106}
\email{cgwinn@condor.physics.ucsb.edu}
\vskip 1 truein
\begin{abstract}
I suggest that 
radio-wave scattering by the
interstellar plasma, in combination with subsonic gradients in the Doppler velocity of
interstellar HI, is responsible for the observed small-scale
variation in HI absorption spectra of pulsars.  
Velocity gradients on the order of 0.05 to $0.3~\kms$ 
across 1~AU can produce the observed variations.
I suggest observational tests to distinguish
between this model and the traditional picture of
small-scale opacity variations from AU-scale cloudlets.
\end{abstract}


\section{INTRODUCTION}

The small-scale structure of cold atomic hydrogen (HI) is an
outstanding problem in the structure of the interstellar medium.
Absorption spectra of the $\lambda=21$~cm hyperfine transition of the ground state
show significant spectra changes over very small angles.  Among the
most extreme of these changes are observed for pulsars \citep{des92,cli88,fra94}.  
Absorption with inferred optical depth of 
$\tau = 0.1$ to $2$ in this line show apparent changes of
$\Delta\tau = 0.01$ to 0.1 over periods of a few months.  Over
this period, the motion of the pulsar typically 
carries the line of sight across
tens of mas, or across several AU, for
absorbing material at a distance of a few hundred pc.  The most
straightforward conclusion is that the HI consists largely of
cloudlets with transverse dimensions of several AU, and optical
depths of 0.01 to 0.1.  This leads to surprisingly dense clouds which
are far from pressure equilibrium with the rest of the interstellar
medium \citep{hei97}.  Small-scale structure of HI 
absorption has also been
observed in spatially-varying absorption of extended continuum sources
\citep{dia89,dav96,fai98,cas00,dha00,fai01}.  
In this paper, I will consider changes in pulsar absorption
quantitatively using a simple model, and discuss absorption
by extended continuum sources only qualitatively.

I propose that the observed variations of
pulsar absorption spectra arise from 
the interaction of interstellar scintillation with transverse gradients in Doppler
velocity of HI. 
Interstellar scintillation is a well-understood phenomenon
resulting from scattering by interstellar free electrons 
\citep[and references therein]{ric77}.  
The observer receives scattered radiation from more than one
path.
Differences
in radio-wave path length, from small-scale fluctuations in the density
of interstellar free electrons, lead to random interference
and so to scintillation.  
At the frequency of the HI line,
absorption and refraction
introduce additional changes in amplitude and phase of the
interfering signals.  
If these additional changes differ among paths,
line and continuum will scintillate differently.
The additional differences among paths 
change rapidly with frequency, across
the line; but only slowly with time.  

I suggest that this differing HI absorption
results from a gradient in Doppler velocity of the HI across the
scattering disk.
At a given observing frequency, absorption by HI
at a range of Doppler velocities in the gradient
will block a strip of radiation,
eliminating some of the paths that would otherwise 
reach the observer and contribute to the interference pattern.
The location of the strip changes with frequency,
blocking different paths and changing the interference pattern
in different ways.
To either side of the blocked strip,
resonant refraction will change the path length and change the
phase along paths.
The consequence is an interference pattern that changes sensitively
with frequency across the absorption line.
This pattern changes with time, at the timescale of scintillation.

In \S\ref{model} of this paper, I present a simple mathematical
model for the effects of a transverse gradient in the Doppler
velocity of HI, or any variation in HI absorption,
in combination with interstellar scattering.
In \S\ref{compare_discuss} I estimate the Doppler velocity gradient
required to produce the observed small-scale variations of HI spectra,
and compare these gradients with those inferred by other observations
and those expected in the absorbing-cloudlet model.
In \S\ref{summary} I summarize the results.

\section{MODEL}\label{model}

\subsection{Scattering by Free Electrons}\label{electron_scattering}

In this section I present a specific model for variations in the HI
line, from sub-AU structure of HI clouds.  Consider a pulsar at
distance $d$ from the Earth, scattered by intervening interstellar
plasma.  In general scattering material will be distributed along the
line of sight; however, a single screen of phase-changing material at
$d/2$ conveniently represents effects of scattering.  The electric
field at the observer can be represented as a Kirchoff integral over
the screen \citep{IO98}.  The observer sees changes in
the intensity of the source because of fortuitous reinforcement or
cancellation of radiation from different points on the scattering
screen.  We suppose that the scattering is
strong: in this case, phase differences between paths are many
radians.  We also assume that the radiation can be treated in the
Gaussian field approximation, in which the phase changes introduced at
different points on the screen are taken to be completely
uncorrelated, so that the summation of electric field 
from different paths at the observer
has the character of a random walk.  However, the argument here
can be generalized to weak or refractive scattering, 
and to power-law or other
spectra for the phase variations over the screen.
In this paper I consider strong diffractive scattering,
for greatest physical insight.

We approximate the Kirchoff integral for electric field at the observer
as a phasor sum:
\begin{equation}
E=\sum_{\imath}{{e^{i \phi_i}}\over{H_{\imath}}} .
\label{efield}
\end{equation}
The phase $\phi_i$ is the sum of that introduced by the scattering screen,
$\Phi(\bx_{\imath})$, and a geometric phase from path length
for the stationary phase point at $\bx_{\imath}$:
$\phi_{\imath}=\Phi(\bx_{\imath})+k x_{\imath}^2 ({4/d})$.
At a stationary phase point,
the gradient of screen phase cancels the gradient of geometric phase.
In the stationary phase approximation 
only stationary phase points contribute to the Kirchoff integral.
The wavenumber is 
$k=2\pi/\lambda$.
The weight of each phasor, $1/H_{\imath}$, corresponds to an area on the screen.
The distribution of stationary phase points, weighted by $1/H_{\imath}$, is 
approximately
a Gaussian distribution, with full width at half maximum $\theta_H d/2$,
in the Gaussian-field approximation.
The size of the region from which the observer receives radiation
acts analogously to a lens, to produce a diffraction pattern
with lateral scale $S_{ISS}\approx \sqrt{4\ln 2}\, \lambda/\theta_H$ in the plane of the observer.

The intensity at the observer is the square modulus of electric field,
\begin{equation}
I_c=E E^* = \sum_{\imath,\jmath} {{e^{i(\phi_{\imath}-\phi_{\jmath})}}\over{H_{\imath} H_{\jmath}}} .
\end{equation}
Here the subscript ``c'' indicates that the intensity is in the continuum,
to distinguish it from line radiation.
The intensity varies with time, with typical timescale $t_{ISS}=S_{ISS}/V_{\perp}$, as the
proper motions of pulsar and screen carry the line of sight through the
scattering material at speed $V_{\perp}$.
The intensity also varies with observing frequency $\nu$,
with typical frequency scale, or ``decorrelation bandwidth,'' 
$\Delta\nu \approx (8\ln 2/ 2\pi)c/(d\theta_H^2)$,
because radiation paths that produce cancellation at one observing 
wavelength may reinforce at another.
\citet{lam99} give precise expressions for these relations for cases of
power-law spectra of density fluctuations.
In strong scattering, decorrelation bandwidth is much smaller than the
observing frequency.  In weak scattering the bandwidth of
scintillations is of order the observing frequency.
Because the electron-density fluctuations responsible for scattering
are dispersive, most pulsars are weakly scattered at sufficiently high
frequencies, and strongly scattered at low frequencies.

Table\ \ref{scatter_obs}
presents the scintillation bandwidths for pulsars with 
observed small-scale structure in the HI line.
For nearly all objects, the decorrelation bandwidth of scintillations is 
somewhat less than the observing
frequency, indicating strong scattering.
However, the decorrelation bandwidth is much greater than
width of the HI line,
of $\approx 3$ to $20~{\rm km\ s}^{-1}$, or $\approx 0.03$ to $0.1$~MHz,
and indeed is much greater than the spectral range typically observed for absorption studies.
The exceptions are
pulsars B0540$+$23 and B1557$-$50.
For B0540$+$23 the decorrelation bandwidth of 0.2 to 0.5 MHz is 
larger than the linewidth, but not larger than the width of a typical spectrum;
\citet{fra94} describe how they remove the variable
baseline introduced by scintillation for this object.
For B1557$-$50 the reported decorrelation bandwidth is much less than the linewidth.
We will consider this object separately.

\subsection{Effects of HI}

At the frequency of the $\lambda=21$~cm line,
intervening HI will 
introduce both absorption and optical path length.
The optical depth $\tau_{HI}$ and the phase change $\phi_{HI}$
parametrize these effects.
Both are proportional to the column density of HI.
The Kramers-Kronig relations relate 
$\tau_{HI}$ and $\phi_{HI}$ to one another
via integrals over all frequencies.
Here we note that, as a rule of thumb,
$\phi_{HI}$ at the line edge, in radians, is
on the order of $\haf \tau_{HI}$ at line center.
The factor of $\haf$ arises from the fact that $\tau$ is
the logarithmic decrease of intensity, rather than electric field.

\begin{figure}[t]
\plotone{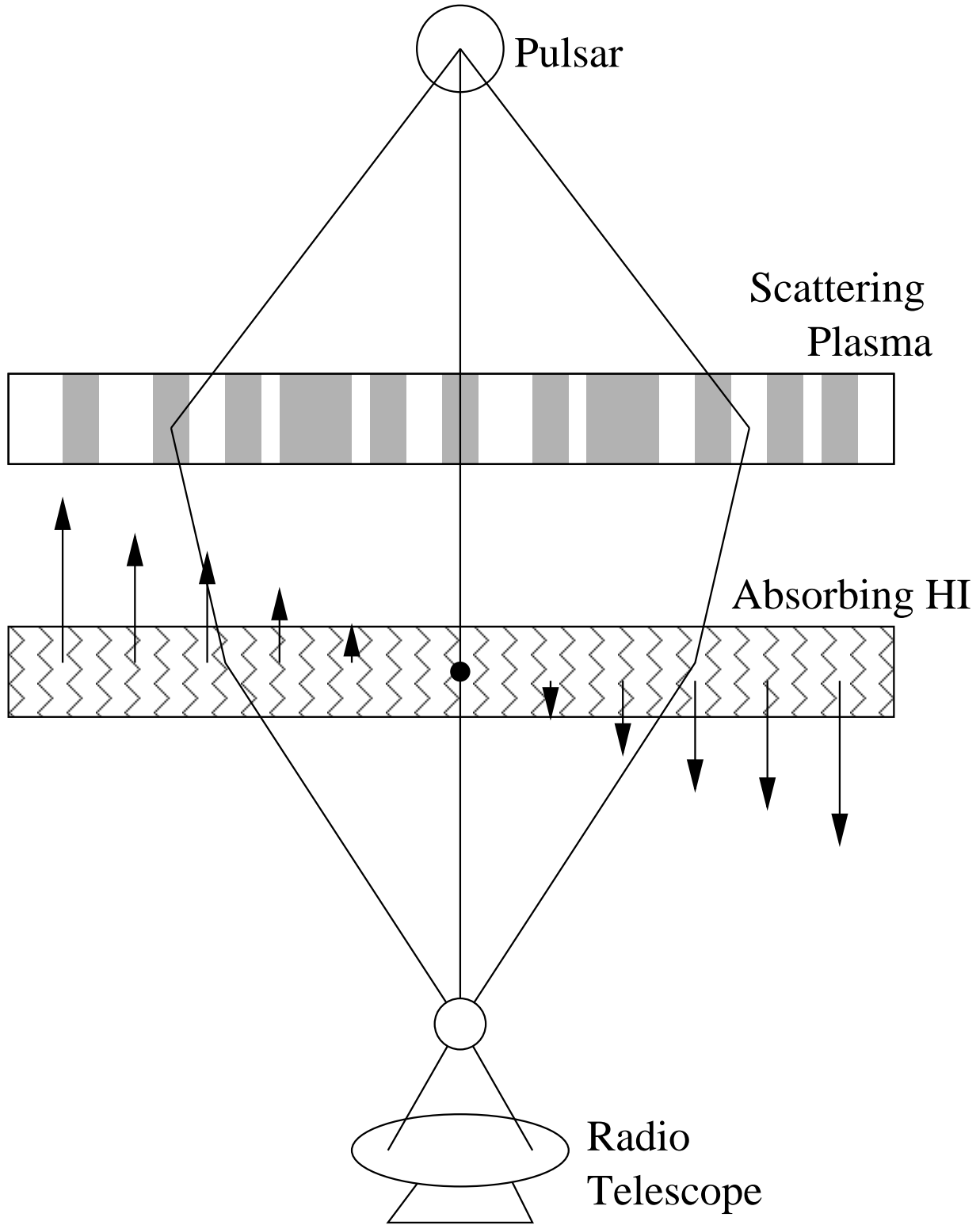}
\figcaption[]{
Schematic view of scattering of
pulsar radiation by fluctuations in the free-electron density of
the interstellar plasma,
followed by absorption by HI
with a transverse gradient of
Doppler velocity.
\label{4planes}} 
\end{figure}

Suppose that the HI consists of a background distribution,
uniform over the screen, and a small fraction that has varying properties over the screen.
Within the absorption line,
the background distribution will introduce a fractional change in amplitude and
a phase rotation, both uniform for all phasors.
The varying part will change each phasor differently
and change the phasor sum $E$.
Figure\ \ref{4planes} shows the geometry schematically,
for the particular case where variation in absorption arises 
from a gradient of Doppler velocity.

A simple mathematical model describes the effects of absorbing
HI in the language of \S\ \ref{electron_scattering}.
Let the constant part have optical depth $\tau_0$ and phase changes of $\varphi_0$,
and the varying part have optical depth $\delta\tau_{\imath}$ and phase change $\delta\varphi_{\imath}$,
at location $\bx_{\imath}$ on the screen.
We suppose that $\delta\tau$ has zero mean, when averaged over the scattering disk,
so that
$\delta\tau_{\imath}$ takes on positive and negative values.
The electric field at the observer, at a frequency within the HI line, is then
a generalization of Eq.\ \ref{efield}:
\begin{equation}
E_{HI}=e^{-\haf\tau_0+i\varphi_0} \sum_{\imath} {{e^{i \phi_i}}\over{H_{\imath}}} e^{-\haf \delta\tau_{\imath}+i \delta\varphi_{\imath}} .
\end{equation}
Through first order in $\delta\tau$ and $\delta\varphi$, the intensity on the line is:
\begin{equation}
I_{HI}=\left|E_{HI}\right|^2 \approx e^{-\tau_0}\left\{ \left|\sum_{\imath}{{e^{i \phi_{\imath}}}\over{H_{\imath}}}\right|^2 
+\re\left[\sum_{\imath}{{e^{i \phi_{\imath}}}\over{H_{\imath}}}\; (\delta\tau_{\imath}+i 2 \delta\varphi_i)\;
\sum_{\jmath}{{e^{-i \phi_{\jmath}}}\over{H_{\jmath}}}\right]
\right\}.
\end{equation}
Studies of the distribution of HI measure the
normalized spectrum.
The normalized absorption of the line is:
\begin{equation}
{{I_c-I_{HI}}\over{I_c}}\approx 1-e^{-\tau_0}
-e^{-\tau_0}\;
\re\left[
{{\sum_{\imath} {\textstyle{e^{i \phi_{\imath}}}\over{H_{\imath}}} \;(\delta\tau_{\imath}+i 2 \delta\varphi_i) } 
\over
{\sum_{\imath} {\textstyle{e^{i \phi_{\imath}}}\over{H_{\imath}}}}} 
\right]
.
\label{fractional_depth}
\end{equation}
The first two terms on the right-hand side give the
optical depth of the constant part of the distribution of HI, as expected.
The last term describes the different scintillation 
on the line and away from the line,
because of the effects of the additional optical depth and phase
of the randomly-varying part of the HI distribution.

The depth of the normalized absorption spectrum, $(I_c-I_{HI})/I_c$,
thus includes a randomly-varying term.
This term is the quotient of 2 random factors.
The variance of the denominator is the mean intensity of the continuum:
$\langle |\sum_i e^{i\phi_i}/H_i|^2\rangle = \langle I_c\rangle$.
Under the assumption that the small-scale fluctuations in electron
density responsible for scattering are uncorrelated with the
varying part of the HI distribution,
the dispersion of the numerator is 
$(\langle |\delta\tau|^2+4\langle |\delta\varphi|^2\rangle)\langle I_c\rangle$.

I assume that the sums in the numerator and denominator have
the statistics of uncorrelated random walks.
This is the case if the Gaussian field approximation holds,
as is the case in strong scattering on short timescales.
(The statistics of HI absorption and refraction are relatively unimportant,
as long as electron-density fluctuations are highly uncorrelated).
For weaker scattering, 
correlations between numerator and denominator can appear:
for example, electron-density fluctuations $\phi$ and 
optical depth can both change linearly with position.
One might expect the variations of the quotient to be even greater
in this case.

The appendix presents the distribution of the quotient of uncorrelated
random walks.
The mean square and higher moments of this distribution do not converge.
However, a ``typical'' value of the quotient
is the quotient of the standard deviations of the distributions,
or $\sqrt{\langle |\delta\tau|^2+4\langle |\delta\varphi|^2\rangle}$
in our case.
Because $\delta\tau$ and $2\delta\varphi$ reach comparable maximum
values (although at different frequencies),
we expect that the quotient will have magnitude 
$\sigma_{\delta\tau}=\langle |\delta\tau|^2\rangle$.

We thus expect variations in intensity across the line
with typical amplitude 
$\Delta I/I_c \approx e^{-\tau_0} \sigma_{\delta\tau}$.
This variation will change amplitude on the timescale
for scintillation, $t_{ISS}$.
On this timescale the sums in the numerator and denominator of
Eq.\ \ref{fractional_depth} will change.
Of course, the intensity of the continuum $I_c$ will
change on the same timescale; however,
these intensity variations have bandwidth much broader
than the observed spectrum, and will be removed by 
averaging in time,
for most pulsars with observed variation of HI spectra.

\subsection{Velocity Gradients of HI}

I suggest that the variations in opacity $\delta\tau$ arise from
velocity gradients.  A cloud with opacity $\tau_c$ and velocity
different $\Delta V$ across length $L$ will show an opacity variation
of $(\Delta V/C) \tau_c$ across that length,
at observing frequencies near
the rest frequencies of the cloud.  
Here $C$ is the thermal velocity in the cloud, $C=\sqrt{k_B T/\mu}$,
where $T$ is the kinetic temperature,
$\mu$ is the mean molecular weight, and $k_B$ is Boltzmann's constant.
We take the opacity difference as $\sigma_{\delta\tau}=(\Delta V/C) \tau_c$
and the length $L$ as the linear size of the scattering disk: $L=\theta_H d/2$.

\begin{figure}[t]
\plotone{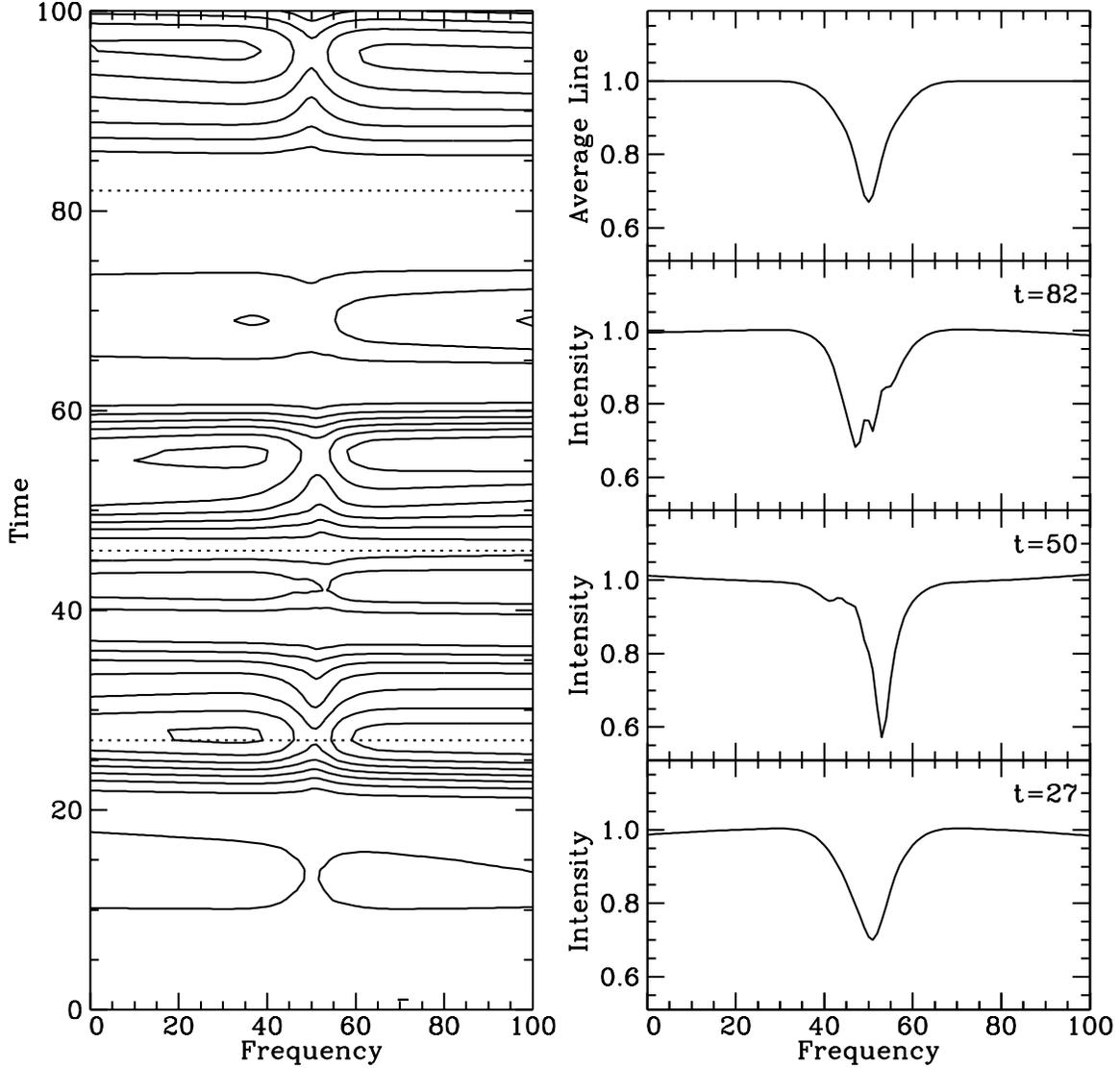}
\figcaption[]{
Simulation of small-scale variation in HI absorption
via gradients in Doppler velocity.
Left: Simulated dynamic spectrum, with HI line at 
center of band.
Right: Upper panel shows average intensity spectrum.
Lower panels show observed spectra at different times,
as indicated by dotted lines on right plot.
A slope has been removed from each spectrum.
Model parameters are average opacity $\tau_0=0.1$, 
velocity gradients 
$\Delta V=0.8 C$ across $\theta_H d/2$
with opacity $\sigma_{\delta\tau}=0.1$,
and total linewidth of 3 times thermal linewidth.
\label{rwalk}} 
\end{figure}

The scintillation will differ at the frequency of the absorption line
and in the continuum.  Both will scintillate,
but not in perfect proportion.
Indeed, because $\delta\tau$ and $\delta\varphi$
vary with frequency,
different parts of the line will scintillate differently.
Figure\ \ref{rwalk} shows a simulation of scintillation
over a frequency range that includes a line with a Gaussian
absorption profile, and associated phase change,
using a phasor sum \citep{IO98}.

\section{COMPARISON AND DISCUSSION}\label{compare_discuss}

\subsection{Comparison with Observations}\label{comparison_w_obs}

Table\ \ref{hi_obs} lists pulsars with observed variations in 
HI absorption.
The table gives the largest $\Delta I/I_c$ observed at each
epoch.
Typical observed variations in intensity are $\Delta I/I_c\approx 0.03$,
and the depths of the normalized absorption spectra
are typically $(I_c-I_{HI})/I_c\approx 0.7$.
For scintillation
to explain this typical case
thus requires $\sigma_{\delta\tau}\approx 0.03/0.7\approx 0.04$.
If all absorption along the line of sight participates
in forming these gradients,
this requires
velocity differences of $\Delta V \approx 0.04 C$
across the diameter of the scattering disk, $L=\theta_H d/2$.
This calculation yields the values displayed in Table\ \ref{theory}.

Among the objects in Table\ \ref{hi_obs},
two show particularly large $\Delta I/I_c$.
These 2 events were the first reported,
and first drew attention to the phenomenon.
Pulsar B$1557-50$ shows $\Delta I/I_c=0.5$ \citep{des92}.
However, this pulsar is by far the most strongly scattered
in the sample, with estimated angular broadening of 121~mas.
The scattering disk diameter is thus 300~AU.
In principle, a weak gradient of $0.003~\kms$ can
produce the result.
However, the scintillation bandwidth is estimated as 160~Hz,
and the scintillation tinescale is probably only a few seconds.
Thus, a single observation averages over many scintillations,
and the scintillation model discussed here is unlikely to
contribute in this particular case.
This object represents an important exception
to the relatively similar scattering properties of
the other pulsars in the sample.

Pulsar B$1821+05$ shows $\Delta I/I_c=0.52$.
The change in the line was 
very narrow in frequency, with a Doppler width of
$<1.2~\kms$, much smaller than the absorption
linewidth of $25~\kms$.
I suggest that this is an unusual event,
perhaps one of the extreme excursions responsible for the
nonconvergence of the mean square of the distribution
of the quotient.

Table\ \ref{theory} shows the gradients in Doppler velocity
required across 1~AU,
for each of the pulsars with observed variability of HI absorption.
For most pulsars I give a range of estimated gradients.
This range includes both the variety of 
$\Delta I/I_c$ reported, and the differences
among measurements of the decorrelation bandwidth.
Because greatest $\Delta I$ across the line is recorded at each
epoch, and because the mean square and higher moments 
of the expected distribution do not
converge, theseXS inferred gradients are more likely to be overestimates
than underestimates.
Lack of accurate measurements of angular broadening also
contributes to the breadth of the range.
Table\ \ref{theory} estimates angular broadening 
from decorrelation bandwidth as
$\theta_H=\sqrt{16\pi\ln 2\; c/\Delta\nu\, d}$.
This estimate
is likely to be rather
poor for nearby objects
such as those in the table \citep{bri98},
although it is accurate to within about a factor of 2
for more strongly-scattered objects \citep{gwi93}.
In summary, interstellar scattering and 
gradients in Doppler velocity of about 0.05 to $0.3~\kms\ {\rm AU}^{-1}$
can produce the observed variations in HI absorption.

\subsection{Observed Gradients in Doppler Velocity}

Although the lengths $L=\theta_H d$ are quite small,
by the standard of typically observed lengths in the interstellar medium,
the inferred velocity differences $\Delta V$ are small as well.
Interstellar clouds invariably support turbulent velocities
of several times thermal velocity.
The widths of interstellar lines are consequently 
greater than those resulting from temperature of the gas
by a factor of 2 to 3, tracing this turbulence \citep{meb82}.
In particular, the 21-cm absorption lines of pulsars
observed for studies of small-scale structure commonly have widths of
2 to 3\ $\kms$, whereas the 50 to 100~K temperature of the gas
would result in Doppler widths of 0.5 to 1\ $\kms$.
These broader linewidths persist at even the highest angular resolutions
observed.

Studies of the distribution of HI absorption against extended
continuum sources may also be affected by scintillation, as discussed
in \S\ \ref{extended_source} below.
Nevertheless, it is interesting to compare velocity
gradients inferred from such studies with those required to
produce small-scale structure of HI absorption for pulsars.
As an example, \citet{fai98} observe a gradient of about $2.5~\kms$
across about 30~mas, toward the source 3C138.
For an absorber distance of 200~pc and a 
thermal velocity of $1~\kms$, this corresponds to a 
gradient of 0.4~$C$/AU, comparable to
the largest of those inferred from the scintillation model,
as shown in Table\ \ref{theory}.

\subsection{Comparison with Cloudlet Model}

Comparison velocity variations in the scattering model
proposed here with density variations required 
in the cloudlet picture is
illuminating.
Under the assumption of rough balance between ram pressure
and thermal pressure,
a velocity difference $\Delta V$ corresponds to a
density difference 
$\Delta\rho\approx \rho V \Delta V/C^2$,
where $\rho$ is the density and $V$ is a typical
random fluid velocity.
The typical random velocity inferred from linewidth
is about the thermal velocity,
so $\Delta\rho\approx \rho \Delta V/C$.
Thus,
in the scintillation model,
the velocity differences of about 0.05 to $0.3~C$
correspond roughly to density differences of
5\% to 30\% across 1~AU.

In the traditional cloudlet model,
with spherical cloudlets,
the required density differences 
are of order a factor of 300 \citep{hei97}.
The discrepancy is ameliorated if HI lies in sheets
or filaments; it is aggravated by the necessity that H$_2$
accompany the HI.
These density differences are expressed across about 30~AU,
the size of the clouds.
In the absence of some additional confining mechanism,
these clouds will evolve over the sound crossing time,
or about 150~yr.
It would be surprising if that evolution did not result in
gradients of the sound speed across 1~AU,
ample to produce variations in intensity by the scattering mechanism
proposed here.
Thus, even if discrete, dense cloudlets exist,
the scattering mechanism proposed here should still play a role.

\subsection{Observational Test}

Small-scale structure of HI absorption
due to scintillation will vary more rapidly 
than structure due to absorbing
cloudlets.
Intensity variations from scintillation vary on the 
scintillation timescale $t_{ISS}=\sqrt{4\ln 2}\, \lambda/(\theta_H V_\perp)$.
The timescale for variations 
from absorbing, intervening cloudlets can be 
no shorter than the time required for the scattering disk 
to move by its own width, $t_r=\theta_H (d/2)/V_{\perp}$,
sometimes called the ``refractive timescale''.
In strong scattering, $t_r$ is much longer than $t_{ISS}$.
Indeed, $t_{ISS}/t_r\approx \Delta\nu/\nu$.
With $\Delta\nu/\nu\approx 0.2$ to 0.01, as shown in Table\ \ref{scatter_obs}
most pulsars studied for small-scale variations
of HI absorption are in strong scattering.
Exceptions are pulsars B0950$+$08,
which is in weak scattering and shows no absorption,
and B1557$-$57, which is far into strong scattering,
as discussed above.
For the other pulsars, observations
of changes in absorption spectra on timescales
$t_{ISS}$ would greatly strengthen the argument
that scintillation, rather than absorbing cloudlets,
is responsible for small-scale structure
of the HI absorption line.

A measurement of
the average intensity of a scintillating source 
must be averaged over about 50 scintillations
to produce a reproducible result,
largely because a few strong scintillations dominate sums
\citep{gwi93}.
Measurements converge even more slowly near the transition
between weak and strong scattering, 
because the fractional modulation is great,
up to $\Delta I/I \approx \sqrt{3}$.
Because most pulsars in the sample have scintillation timescales
of a few thousand seconds,
even observations of several hours can fail
to eliminate effects of scintillation by time averaging.

Pulsar 0540$+$23 is an exception.
It has a scintillation timescale short enough that 
spectra from several contiguous scintillation timescales could
be observed and compared,
and much shorter than the refractive timescale.
Thus the timescale 
of variations of HI absorption for this object
presents an important test of the origin of variations in
HI absorption.

\subsection{Structure on a Range of Scales}\label{powerlaw}

\citet{des00} points out that
absorbing HI is likely to be distributed on a variety of
length scales.  As an extreme case, a large-scale gradient of optical depth
will produce spectral variations over any scale large
enough not to be smeared by scattering.
In this case, the smallest scale showing spectral variations has nothing
to do with the size of the cloud.
A medium with structure on a variety of scales is most accurately
described by a structure function, or equivalently, spatial spectrum.

If velocity gradients in HI combine with interstellar scintillation
to produce small-scale structure of the HI absorption line,
the structure function of that structure will reflect
different physics at large and small scales.
Effects of scintillation at the short length scale
$V_{\perp} t_{ISS} << 1$~AU and do not increase 
with further increases inlength scale.
Absorption by discrete clouds
Will become increasingly important with
increasing scale, as effects of larger clouds
become more important.
Determination of the structure function,
and the point at which the underlying physics changes,
will help to understand interstellar HI as well as
the origin of small-scale changes in absorption spectra.

\subsection{Extended Continuum Sources and Weak Scattering}\label{extended_source}

As noted above, small-scale variations of HI absorption
have been observed toward several continuum sources 
\citep{dia89,dav96,fai98,dha00,fai01}.
Such sources do not scintillate,
because they are large enough to quench scintillation.
More precisely, 
the source is large enough that different parts scintillate
independently and cancel out.

I suggest scintillation
could nevertheless explain the observed variations.
Many of the sources may be in weak scintillation,
where the maximum source size for scintillation is relatively large,
and scintillation is quenched only slowly as source size increases.
Moreover, the size relevant as a constraint,
for resolved sources in weak scattering, 
is the resolution of the instrument rather than
the size of the source.
Analogously, stars twinkle, but planets do not;
but a single point on a planet's surface does twinkle.
Thus, a single point on an extended radio source
will show variations in HI absorption,
if scattering by electron-density fluctuations and a 
gradient of Doppler velocity of HI are present.

Refractive scattering can also produce variations in surface
brightness of extended sources \citep{ng89,gn89,mu90}.
This refractive ``mottling'' of the source
is not quenched by finite source size.
Velocity gradients will not affect refractive scattering,
unless they are very large; however,
higher-order polynomial variations can affect it.
Although the extension of the formalism of this paper to
extended sources is straightforward in principle,
description of weak and refractive 
scintillation requires a more complicated
formalism, and I defer it to a subsequent paper.

\section{SUMMARY}\label{summary}

I have presented a simple model for the effects of absorption
by HI with a spatially-varying opacity $\delta\tau$ on
a scintillating pulsar.
The scintillation state is different on and off the line:
both scintillate, but not synchronously.
The bandwidth of scintillation of the continuum is 
set by the phase differences among different paths,
and is much broader than the line;
the bandwidth of scintillation on the line is set by
variations in opacity, and is typically
of order the thermal linewidth.

Gradients in Doppler velocity of the absorbing HI
can produce relative
scintillations of line and continuum.
Gradients 
of order 0.05 to 0.3 times the thermal velocity, across 1~AU
produce effects
effects comparable to the observed
small-scale variations of HI absorption spectra.
Such gradients are comparable to those inferred from observations.
If discrete absorbing cloudlets produced the variation,
they would result in comparable or larger gradients.
The timescale of scintillation, on and off the line,
is equal to the scintillation timescale,
and can be much shorter than the minimum time
for variations in intensity from absorbing cloudlets.
This difference in timescales,
and difference in structure functions,
present possible observational tests of the scintillation model 
presented here.

\acknowledgments

I thank A. Deshpande and J. Heyl for useful discussions,
and J. Weisberg for a copy of his thesis.
The U.S. National Science Foundation provided financial support.

\appendix
\section{QUOTIENT OF ELEMENTS DRAWN FROM UNCORRELATED GAUSSIAN DISTRIBUTIONS}\label{quotient}

Consider the distribution of $f=u/x$,
where $u$ and $x$ are complex variables
drawn from uncorrelated complex Gaussian distributions:
\begin{eqnarray}
P(x)&=&{{1}\over{2\pi \sigma_x^2}} e^{-\haf |x|^2/\sigma_x^2} \\
P(u)&=&{{1}\over{2\pi \sigma_u^2}} e^{-\haf |u|^2/\sigma_u^2} . \nonumber
\end{eqnarray}
In this case, the distribution of $f$ is:
\begin{equation}
P(f)=\int du \int dx P(u) P(x)\; \delta(\, u/x-f),
\end{equation}
where $\delta$ is the Dirac delta-function.  I define
\begin{eqnarray}
R_u=\re[u],&\quad &  I_u=\im[u],\\
R_x=\re[x],&\quad &  I_x=\im[x],\nonumber \\
R_f=\re[f],&\quad &  I_f=\im[f].\nonumber 
\end{eqnarray}
Now one finds:
\begin{eqnarray}
P(f)&=&{{1}\over{4\pi^2 \sigma_u^2 \sigma_x^2}}
\int du \int dx\,  e^{-\haf {{|u|^2}\over{\sigma_u^2}}}\, e^{-\haf {{|x|^2}\over{\sigma_x^2}}}\, 
\delta (\re[u/x]-R_f) \delta (\im[u/x]-I_f) \\
&=&{{1}\over{4\pi^2 \sigma_u^2 \sigma_x^2}}
\int du \int dx  e^{-\haf {{|x|^2}\over{\sigma_x^2}}}\, e^{-\haf {{|u|^2}\over{\sigma_u^2}}} \,
\delta \left({{R_u R_x - I_u I_x}\over{R_x^2+I_x^2}}-R_f\right) \,
\delta \left({{I_u R_x - R_u I_x}\over{R_x^2+I_x^2}}-I_f\right). \nonumber 
\end{eqnarray}
Note that:
\begin{equation}
|u|^2=|f|^2 |x|^2 
\end{equation}
and
\begin{equation}
|x|^2 = R_x^2 + I_x^2 .
\end{equation}
Using the rule $\delta(x)=|a| \delta( a x)$,
one can then transform the $\delta$-functions,
and the argument of the exponential in $|u|^2$,
to find:
\begin{equation}
P(f)={{1}\over{4\pi^2 \sigma_u^2 \sigma_x^2}}
\int du \int dx\; e^{-\haf({{1}\over{\sigma_x^2}}+{{|f|^2}\over{\sigma_u^2}}) |x|^2 }\;
{{|x|^2}\over{I_x}}\, \delta\left(I_u-{{R_f |x|^2}\over{I_x}}+{{R_x R_u}\over{I_x}}\right)
I_x \delta(R_u-R_f R_x + I_f I_x) ,
\end{equation}
where I have used the transformed first $\delta$-function to
transform the second.
The integrations over $u$ and the complex phase of $x$ are now trivial,
and the distribution is:
\begin{equation}
P(f)={{1}\over{2\pi \sigma_u^2 \sigma_x^2}}\; \int_0^{\infty}\; dX\; X^3\, 
\exp\left\{-\haf \left({{1}\over{\sigma_x^2}}+{{|f|^2}\over{\sigma_u^2}}\right) X^2 \right\},
\end{equation}
where $X=|x|$.
This can be evaluated to find
\begin{equation}
P(f)={{\sigma_x^2 \sigma_u^2}\over{\pi (\sigma_u^2 + |f|^2 \sigma_x^2)^2 }}.
\end{equation}
Note that the distribution function is normalized to unit
volume over the complex plane, and
that it is independent of the complex phase of $f$.
This is a consequence of the fact that $u$ and $x$ are uncorrelated.

We can find the distribution of $R_f=\re[f]$ 
by integrating $P(f)$ over $\im[f]$.
We find:
\begin{equation}
P(R_f)={{\sigma_x \sigma_u^2 }\over{2 (\sigma_u^2 + R_f^2 \sigma_x^2)^{3/2}}} .
\end{equation}
This distribution is normalized to unit area, over the real axis. 
It is symmetric about 0 and all odd moments of $P(R_f)$ vanish.
The mean of the absolute value of $R_f$ is $\langle |R_f|\rangle = 1$.
However, $\langle R_f^2\rangle$ and higher moments do not converge.
Thus, one cannot speak of a mean squared value of $R_f$,
and one cannot use this traditional measure to characterize the distribution.
However, one can describe alternative measures.
For example, a fraction $P_0$ of the full distribution
will lie between
the values $-f_0$ and $+f_0$ when:
\begin{equation}
f_0={{\sigma_u P_0}\over{\sigma_x \sqrt{1-P_0^2}}} .
\end{equation}
Thus, 50\% of time time the real part of the 
quotient $R_f$ will lie at less than $f_{50}=1/\sqrt{3}\,\sigma_u/\sigma_x$,
75\% of the time $R_f$ will lie at less than $f_{75}=\sqrt{9/7}\,\sigma_u/\sigma_x \approx 1.1\, \sigma_u/\sigma_x$,
and 95\% of the time $R_f$ will lie below $f_{95}=\sqrt{391/39}\,\sigma_u/\sigma_x \approx 3\, \sigma_u/\sigma_x$.

\newpage

\begin{deluxetable}{lccr}
\tablewidth{300pt}
\tablecaption{Scattering Properties of Pulsars}
\tablehead{
\colhead{      }&\colhead{Decorrelation}&\colhead{Scintillation}&\colhead{          }\\
\colhead{Pulsar}&\colhead{Bandwidth\tablenotemark{a}}&\colhead{Timescale\tablenotemark{b}}&\colhead{References}\\
\colhead{      }&\colhead{$\Delta\nu$  }&\colhead{$t_{ISS}$    }&\colhead{          }\\
\colhead{      }&\colhead{(MHz)        }&\colhead{(sec)        }&\colhead{          }
}
\startdata
B0540$+$23& 0.2~to~ 0.5& 58          & 1,2 \\
B0823$+$26& 160~to~ 180& 550~to~940  & 2,3 \\
B0950$+$08& weak       & weak        & 4   \\
B1133$+$16& 280~to~ 900& 390~to~800  & 2,3 \\
B1557$-$50& 0.00016    &             & 5   \\
B1737$+$13& 5~to~8     &             & 1   \\
B1821$+$05& 3.7~to~4.6 &         1500& 1,2 \\
B1929$+$10& 770~to~648 & 1500~to~1900& 2,3 \\
B2016$+$28& 30~to~130  & 4400~to~2500& 2,3 \\
\enddata
\tablenotetext{a} {Scaled to frequency $\nu=1420$~MHz via $\Delta\nu\propto \nu^{4.4}$.  Ranges indicate range of observed values.}
\tablenotetext{b} {Scaled to frequency $\nu=1420$~MHz via $t_{ISS}\propto \nu^{1.2}$.  Ranges indicate range of observed values.}
\tablerefs{
(1) \citet{cor85};
(2) \citet{gup95};
(3) \citet{bha99};
(4) \citet{phi92};
(5) \citet{ric77}.}
\label{scatter_obs}
\end{deluxetable}


\begin{deluxetable}{lclclr}
\tablewidth{400pt}
\tablecaption{Reported Variations of HI Absorption}
\tablehead{
\colhead{      }&\colhead{      }&\colhead{Time}&\colhead{Normalized }&\colhead{Fractional      }&\colhead{          }\\
\colhead{Pulsar}&\colhead{Epochs}&\colhead{Span}&\colhead{Absorption }&\colhead{Intensity Change\tablenotemark{a}}&\colhead{References}\\
\colhead{      }&\colhead{      }&\colhead{    }&\colhead{$(I_c-I_{HI})/I_c$}&\colhead{$\Delta I_{HI}/I_c$  }&\colhead{          }\\
\colhead{      }&\colhead{      }&\colhead{(yr)}&\colhead{                  }&\colhead{                }&\colhead{          }
}
\startdata
B0540$+$23& 3 & 1.7 & 0.90   & 0.06,  0.08,  0.05  & 1 \\
B0823$+$26& 3 & 1.7 & 0.75   & 0.025, 0.045, 0.045 & 1 \\
B0950$+$08& 3 & 1.7 & 0\phantom{.00}      & 0                   & 1 \\
B1133$+$16& 3 & 1.7 & 0.75   & 0.013, 0.043, 0.022 & 1 \\
B1557$-$50& 2 & 4   & 0.70   & 0.5                 & 2 \\
B1737$+$13& 3 & 1.7 & 0.27   & 0.09,  0.06,  0.06  & 1 \\
B1821$+$05& 2 & 1   & 0.68   & 0.52                & 3,4 \\
B1929$+$10& 3 & 1.7 & 0.88   & 0.02,  0.009, 0.015 & 1 \\
B2016$+$28& 4 & 4.8 & 0.21   & 0.08,  0.03,  0.05  & 1 \\
\enddata
\tablenotetext{a} {Multiple entries reflect multiple epochs.}
\tablerefs{
(1) \citet{fra94};
(2) \citet{des92};
(3) \citet{cli88};
(4) \citet{fra91}. }
\label{hi_obs}
\end{deluxetable}


\newpage

\begin{deluxetable}{lcccc}
\tablewidth{350pt}
\tablecaption{Inferred Velocity Gradient and Scale}
\tablehead{
\colhead{      }&\colhead{        }&\colhead{Angular   }&\colhead{Length          }&\colhead{Velocity              }\\           
\colhead{Pulsar}&\colhead{Distance}&\colhead{Broadening\tablenotemark{a}}&\colhead{Scale           }&\colhead{Gradient              }\\
\colhead{      }&\colhead{$d$     }&\colhead{$\theta_H$}&\colhead{$L=\theta_H d/2$}&\colhead{$\Delta V/(C L)$    }\\
\colhead{      }&\colhead{(kpc)   }&\colhead{(mas)     }&\colhead{(AU)            }&\colhead{($\kms~{\rm AU}^{-1}$)}
}
\startdata
B0540$+$23 &   3.54 &	 3   to	5    &	  5    to  9    &	0.015 to 0.04 \\ 
B0823$+$26 &   0.38 &	 0.5  	     &	  0.3  to       &	0.15 to 0.3   \\ 
B1133$+$16 &   0.27 &	 0.2 to	0.4  &	  0.12 to  0.2  &	0.10 to 0.6   \\ 
B1557$-$50\tablenotemark{b} &   6.30 &  121   	     &	303             &	0.003         \\ 
B1737$+$13 &   4.77 &	 0.6 to	0.8  &	  1.3  to  1.7  &	0.04 to 0.08  \\ 
B1821$+$05\tablenotemark{c} &   3.0  &	 1.0 to	1.2  &	  1.8  to  2.0  &	0.4           \\ 
B1929$+$10 &   0.17 &	 0.3 to	0.4  &	  0.14 to  0.15 &	0.14 to 0.3   \\ 
B2016$+$28 &   1.10 &	 0.3 to	0.7  &	  0.3  to  0.7  &	0.05 to 0.3   \\ 
\enddata
\tablenotetext{a} {Estimated from decorrelation bandwidth and distance as 
$\theta_H\ =\ \sqrt{16 \pi \ln 2 c/ \Delta\nu\, d}$ \citep{gwi93}.}
\tablenotetext{b} {Long period between observations; variations may reflect changes in $\tau_0$.}
\tablenotetext{c} {Large $\Delta I/I_c$ is responsible for large gradient.}
\tablecomments{Parameter ranges indicate the observed ranges of $\Delta\nu$ and $\Delta I_{HI}/I_c$ in
Table 1, and should be taken to represent the range in which the derived parameter is likely to lie,
rather than an observational uncertainty, or the true range of variation of the parameter.}
\label{theory}
\end{deluxetable}

\newpage
\figurenum{1}

\newpage
\figurenum{2}


\clearpage

\begin{thebibliography}{}

\bibitem[Bhat et al.(1999)]{bha99} Bhat, N.D.R., Rao, A.P., \& Gupta, Y. 1999, ApJS, 121, 483 

\bibitem[Britton, Gwinn, \& Ojeda(1998)]{bri98}Britton, M.C., Gwinn, C.R., \& Ojeda, M.J. 1998, ApJ, 501, L101 

\bibitem[Clifton et al.(1988)]{cli88}Clifton, T.R., Frail, D.A., Kulkarni, S.R., \& Weisberg, J.M. 1988, ApJ, 333, 332 

\bibitem[Cordes, Weisberg \& Boriakoff(1985)]{cor85} Cordes, J.M., Weisberg, J.M., \& Boriakoff, V. 1985, ApJ, 288, 221

\bibitem[Davis, Diamond, \& Goss(1996)]{dav96}Davis, R. J., Diamond, P. J., \& Goss, W. M. 1996, MNRAS, 283, 1105

\bibitem[Deshpande(2000)]{des00} Deshpande, A.A. 2000, MNRAS, 317, 199 

\bibitem[Deshpande(1992)]{cas00}Deshpande, A.A., Dwarakanath, K.S., \& Goss, W.M. 2000, ApJ, 543, 227 

\bibitem[Deshpande et al.(1992)]{des92} Deshpande, A.A., McCulloch, P.M., Radhakrishnan, V., \& Anantharamaiah, K.R. 1992, MNRAS, 258, 19P 

\bibitem[Dhawan, Goss, \& Rodriguez(2000)]{dha00} Dhawan, V., Goss, W.M., \& Rodriguez, L.F. 2000, ApJ, 540, 863 

\bibitem[Diamond et al.(1989)]{dia89} Diamond, P. J., Goss, W. M., Romney, J. D., Booth, R. S., Kalberla, P. M. W., \& Mebold, U. 1989, ApJ, 347, 302

\bibitem[Faison et al.(1998)]{fai98}Faison, M.D., Goss, W.M., Diamond, P.J., \& Taylor, G.B. 1998, AJ, 116, 2916 

\bibitem[Faison \& Goss(2001)]{fai01}Faison, M.D., \& Goss, W.M. 2001, AJ, in press

\bibitem[Frail et al.(1991)]{fra91}Frail, D.A., Cordes, J.M., Hankins, T.H., \& Weisberg, J.M. 1991, ApJ, 382, 168 

\bibitem[Frail et al.(1994)]{fra94}Frail, D.A., Weisberg, J.M., Cordes, J.M., \& Mathers, C. 1994, ApJ, 436, 144 

\bibitem[Goodman \& Narayan(1989)]{gn89}Goodman, J., \& Narayan, R. 1989, MNRAS, 238, 995

\bibitem[Gupta(1995)]{gup95} Gupta, Y. 1995, ApJ, 451, 717 

\bibitem[Gwinn, Bartel, \& Cordes(1993)]{gwi93}Gwinn, C.R., Bartel, N., Cordes J.M. 1993, ApJ, 410, 673 

\bibitem[Gwinn et al.(1998)]{IO98}Gwinn, C.R., Britton, M.C., Reynolds, J.E., Jauncey, D.L., King, E.A., McCulloch, P.M., Lovell, J.E.J., \& Preston, R.A. 1998, ApJ, 505, 928 

\bibitem[Heiles(1997)]{hei97}Heiles, C. 1997, ApJ, 481, 193 

\bibitem[Lambert \& Rickett(1999)]{lam99}Lambert, H.C., \& Rickett, B.J. 1999, ApJ, 517, 299 

\bibitem[Mebold et al.(1982)]{meb82}Mebold, U., Winnberg, A., Kalberla, P.M.W., \& Goss, W.M. 1982, A\&A, 115, 223

\bibitem[Mutel \& Lestrade(1990)]{mu90}Mutel, R.L., \& Lestrade, J.-F. 1990, \apj, 349, L47 

\bibitem[Narayan \& Goodman(1989)]{ng89}Narayan, R. \& Goodman, J. 1989, MNRAS, 238, 963

\bibitem[Phillips \& Clegg(1992)]{phi92}Phillips, J.A., \& Clegg, A.W. 1992, Nature, 360, 137 

\bibitem[Rickett(1977)]{ric77}Rickett, B.J., 1977, ARA\&A, 15, 479


\end{thebibliography}
\end{document}